# Inhomogeneous charge ordering of a spinless fermionic system on the Bethe Lattice

Ferdinando Mancini[1], Francesco Paolo Mancini[2], Adele Naddeo[1]

1 Dipartimento di Fisica "E. R. Caianiello"- Unità CNISM di Salerno, Università di Salerno, 84081 Baronissi (SA), Italy
2 Dipartimento di Fisica and Sezione INFN, Università di Perugia, 06123 Perugia, Italy

**Abstract**
We show that a system of spinless Fermi particles, localized on the sites of the Bethe lattice with coordination number $z$ and interacting through a repulsive nearest-neighbor interaction, exhibits a phase transition to a charge-ordered state. The phase diagram in the $n$-$T$ plane is derived. Relevant thermodynamic quantities, such as the free energy, the specific heat, the entropy and the compressibility are analyzed in detail.

**1. Introduction**

A system of spinless Fermi particles, localized on the sites of the Bethe lattice with coordination number $z$ and interacting through an attractive nearest-neighbor interaction $V$, has been exactly solved [1] by means of the equations of motion approach [2]. This system is shown to be isomorphic to a spin-1/2 ferromagnetic Ising model in an external magnetic field. A complete set of eigenoperators has been found together with the corresponding eigenvalues. Then, Green's and correlation functions have been determined and the relevant thermodynamic quantities have been studied in detail. The phase diagram has been obtained by fixing the particle density $n$, the nearest-neighbor interaction $V$ and the temperature $T$ as external thermodynamic parameters and allowing the system to respond by adjusting the chemical potential $\mu$ [1]. We found the existence of a critical temperature below which there is a spontaneous breakdown of the particle-hole symmetry and the system is unstable against the formation of inhomogeneous phases.

In this paper, we study the case of repulsive nearest-neighbor interaction $V$ and show that a phase transition to a charge ordered state takes place. This case corresponds to the spin-1/2 Ising model in a magnetic field with an antiferromagnetic coupling. This system has already been investigated on the Bethe lattice with coordination number $z$ [3-8] and an exact solution has been found, whose equivalence to the Bethe approximation has also been shown. In all these approaches statistical methods, such as the matrix transfer method and iteration techniques, have been used. In the fermionic version of the model we use quantum field techniques [9], based on the use of the equation of motion method and algebraic properties of the field operators. By using such a formalism, in a series of works [9] we have shown that there is a large class of fermionic systems for which it is possible to find, for any dimension, a finite closed set of eigenoperators and eigenvalues of the Hamiltonian. Then, the hierarchy of the equations of motion closes and analytical exact expressions for the Green's functions are obtained in terms of a finite number of parameters, to be self-consistently determined.

By using this formalism, we study the model by looking for a solution where the translational invariance is spontaneously broken and a charge ordered state, thermodinamically stable, is formed below a critical temperature. We obtain the phase diagram in the $n$-$T$ plane. Relevant thermodynamic quantities, such as the free energy, the specific heat, the entropy and the compressibility are analyzed in detail focusing, for the sake of simplicity, on the particular case $z=3$.

**2. Model Hamiltonian and inhomogeneous solution**

Let us consider a system of spinless fermionic fields, localized on a lattice and interacting by a two-body intersite potential. The corresponding Hamiltonian is given by:

$$H = -\mu \sum_{\mathbf{i}} n(i) + \frac{1}{2} \sum_{\mathbf{i} \neq \mathbf{j}} V_{\mathbf{ij}} n(i) n(j) \,, \qquad (2.1)$$

where $n(i) = c^\dagger(i) c(i)$ is the charge density operator, $c(i)$ ($c^\dagger(i)$) is the fermionic annihilation (creation) operator, satisfying canonical anti-commutation relations. We use the Heisenberg picture: $i = (\mathbf{i}, t)$, where $\mathbf{i}$ stays for the lattice vector $\mathbf{R}_i$. $\mu$ is the chemical potential. We shall study this model on the Bethe lattice with coordination number $z$ and consider only nearest neighbor interactions. Then, Hamiltonian (2.1) can be written as

$$H = -\mu \sum_i n(i) + \frac{1}{2} z V \sum_i n(i) n^\alpha(i) \,, \qquad (2.2)$$

where

$$n^\alpha(i) = \frac{1}{z} \sum_{p=1}^{z} n(i_p) \,, \qquad (2.3)$$

$i_p$ ($p = 1, \cdots z$) being the nearest neighbor of site $i$. We shall solve the Hamiltonian (2.2) with the following boundary conditions:

$$\langle n(i) \rangle = \begin{cases} n_1 & i \in even\ shell \\ n_2 & i \in odd\ shell \end{cases} \qquad (2.4)$$

$$n = \frac{1}{N}\sum_i <n(i)> = \frac{1}{2}(n_1 + n_2), \tag{2.5}$$

that is, we shall investigate the existence of an inhomogeneous solution. Here $N$ is the number of the sites. To such an extent, let us take two distinct sites $i$ and $j$, and consider the two representations

$$H = H_0^{(a)} + H_I^{(a)}$$
$$H_I^{(a)} = zVn(i)n^\alpha(i) \qquad (a = i, j). \tag{2.6}$$

Let us observe that for any operator $O$, the thermal average $<O> = Tr\{Oe^{-\beta H}\}/Tr\{e^{-\beta H}\}$ can be expressed as

$$<O> = \frac{<Oe^{-\beta H_I^{(a)}}>_a}{<e^{-\beta H_I^{(a)}}>_a}, \tag{2.7}$$

where the symbol $<\cdots>_a$ stands for the trace with respect to the reduced Hamiltonian $H_0^{(a)}$

$$<\cdots>_a = \frac{Tr\{\cdots e^{-\beta H_0^{(a)}}\}}{Tr\{e^{-\beta H_0^{(a)}}\}} \qquad (a = i, j). \tag{2.8}$$

By using the algebraic property of the particle operator $[n(i)]^m = n(i)$, we can write

$$e^{-\beta H_I^{(a)}} = \prod_{p=1}^{z} [1 + An(a)n(a_p)], \tag{2.9}$$

where $A = e^{-\beta V} - 1$. Then, by using the property of the $H_0^{(a)}$-representation, we obtain:

$$<e^{-\beta H_I^{(a)}}>_{(a)} = 1 + [(1 + AX_a)^z - 1] <n(a)>_a \tag{2.10}$$

where we put:

$$X_a = <n(a_1)>_a = <n(a_2)>_a = \cdots = <n(a_z)>_a. \tag{2.11}$$

In the $H_0^{(a)}$-representation $c(a)$ satisfies the equation of motion

$$i\frac{\partial}{\partial t}c(a) = -\mu c(a), \tag{2.12}$$

which immediately tell us that

$$<n(a)>_a = \frac{1}{e^{-\beta\mu} + 1}. \tag{2.13}$$

By noting that

$$<n(a)e^{-\beta H_I^{(a)}}>_a = <n(a)>_a (1 + AX_a)^z, \tag{2.14}$$

$$<n(a_p)e^{-\beta H_I^{(a)}}>_a = X_a(1 - <n(a)>_a) + (1 + A)<n(a)>_a X_a[1 + AX_a]^{z-1}, \tag{2.15}$$

we have

$$<n(a)> = \frac{<n(a)>_a (1 + AX_a)^z}{1 + [(1 + AX_a)^z - 1]<n(a)>_a}, \tag{2.16}$$

$$<n(a_p)> = \frac{X_a(1 - <n(a)>_a) + (1 + A)<n(a)>_a X_a[1 + AX_a]^{z-1}}{1 + [(1 + AX_a)^z - 1]<n(a)>_a}. \tag{2.17}$$

By taking $j = i_p$ and imposing the boundary conditions (2.4), we obtain the following equations:

$$(1 + AX)^z[e^{-\beta\mu} + (1 + AY)^z] = [e^{-\beta\mu} + (1 + AX)^z]Y[e^{-\beta\mu} + (1 + A)(1 + AY)^{z-1}]$$
$$(1 + AY)^z[e^{-\beta\mu} + (1 + AX)^z] = [e^{-\beta\mu} + (1 + AY)^z]X[e^{-\beta\mu} + (1 + A)(1 + AX)^{z-1}], \tag{2.18}$$

where for simplicity we have define $X = X_i$ and $Y = X_j$. Numerical and analytical studies show that these equations are redundant. By eliminating unphysical (i.e., complex and divergent) solutions, Eqs. (2.18) take the simpler form

$$(1 - Y)(1 + AX)^{z-1} = e^{-\beta\mu}Y$$
$$(1 - X)(1 + AY)^{z-1} = e^{-\beta\mu}X. \tag{2.19}$$

These equations will fix the two parameters $X$ and $Y$, while the values of the particle density are given by:

$$n_1 = <n(i)> = \frac{(1 + AX)^z}{e^{-\beta\mu} + (1 + AX)^z}$$
$$n_2 = <n(i_p)> = \frac{(1 + AY)^z}{e^{-\beta\mu} + (1 + AY)^z}. \tag{2.20}$$

Equations similar to (2.19) have been previously derived in the context of the spin-1/2 Ising model [3]. The correspondence between the notations is:

$$X = \frac{e^{-2\beta J}}{l_\alpha + e^{-2\beta J}} \qquad Y = \frac{e^{-2\beta J}}{l_\beta + e^{-2\beta J}} \qquad (2.21)$$

$$\mu = 2(h - zJ) \qquad V = -4J.$$

It is worthwhile to notice that the parameters $l_\alpha$ and $l_\beta$, introduced in Ref. [3], do not have a direct physical meaning, whereas the parameters $X$ and $Y$ are particle density expectation values in the $H_0^{(a)}$-representation.

The set of equations (2.19) contains as unknown parameter the chemical potential $\mu$. This quantity can be determined by means of the self-consistent equation

$$n = \frac{1}{2}(n_1 + n_2), \qquad (2.22)$$

where $n_1$ and $n_2$ can be computed by means of Eq. (2.20). As a result, we have a set of three coupled equations, (2.19) and (2.22), which will fix the internal parameters $X$, $Y$ and $\mu$ in terms of the external parameters $z$, $n$, $T$ and $V$. Once the internal parameters are known, all the physical quantities of the system can be calculated, as it will be shown in the next Sections.

## 3. Study of the phase diagram

In this Section we shall look for solutions of Eqs. (2.19) in order to study the phase diagram of the system and then the relevant thermodynamic quantities. We shall focus on the $z=3$ case and show how non-trivial physical consequences arise already in such a simple case. For a general value of $\mu$, let us notice that the set of equations (2.19) can be factorized as:

$$\begin{cases} X = Y \\ (1-X)(1+AX)^2 - GX = 0 \end{cases} \qquad (3.1)$$

$$\begin{cases} Y = \dfrac{1 + G + A(2 + AX)}{A^2(X-1)} \\ aX^2 + bX + c = 0 \end{cases} \qquad (3.2)$$

where $G = e^{-\beta\mu}$ and we have defined

$$a = 2A^2\left[(1+A)^2 + G\right]$$
$$b = 2A\left[2(1+A)^2 + (2+A)G\right] \qquad (3.3)$$
$$c = 2(1+A+G)^2.$$

The first set (3.1) corresponds to the homogeneous solution (i.e., $<n(i)>$ does not depend on the site) and has been already studied in our previous works [1]. Therefore, we shall concentrate the attention on the second set (3.2), which gives the following solution:

$$\begin{cases} X = \dfrac{-b - \sqrt{b^2 - 4ac}}{2a} \\ Y = \dfrac{-b + \sqrt{b^2 - 4ac}}{2a}. \end{cases} \qquad (3.4)$$

Let us concentrate on the quantity $Q = b^2 - 4ac$, which depends on the parameters $V$, $T$ and $\mu$. When $V<0$, we find that $Q$ is always negative; that is, for attractive intersite interaction there is no inhomogeneous phase. On the other hand, for $V>0$ there exists a region in the parameter space where $Q>0$ and, thus, an inhomogeneous solution does exist. A numerical study of the condition $Q>0$ gives the phase diagram of the system. We would like to stress here that the external thermodynamic parameters are $n$ and $T$ (we take as unit of energy $V=1$) and the system responses to their variation by adjusting the chemical potential. Since $n$ and $-\mu$ are conjugated variables, it is physically meaningful to fix $n$ and then determine $\mu$ by means of the self-consistent equation (2.22). A numerical study of the condition $Q>0$ gives the phase diagram in the $(n,T)$-plane, as shown in Fig. 1.

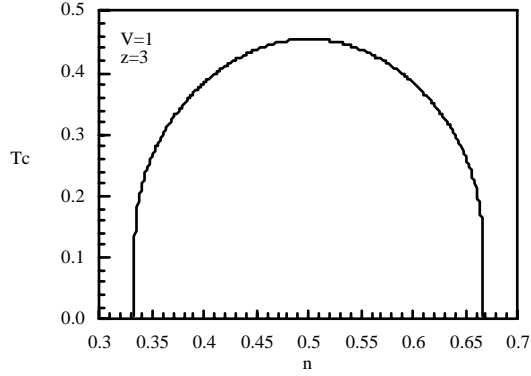

Fig. 1 Phase Diagram in the (*n*,*T*)-plane.

Let us observe that at zero temperature the critical values of *n* are 1/3 and 2/3. Below the curve the system is stable and a charge ordered state is established, while for $T > T_c$ a transition to the homogeneous phase takes place. This result is well confirmed by the behavior of different physical quantities computed in the next Section.

## 4. Thermodynamic quantities

The aim of this Section is to study the relevant thermodynamic quantities. Let us start from the internal energy per site, which has the expression:

$$E(T) = <H> = \frac{1}{2N} zV \sum_i <n(i)n^\alpha(i)> .  \quad (4.1)$$

The correlation function $<n(i)n^\alpha(i)>$ is obviously independent on the site and can be calculated by means of Eqs. (2.7), (2.9) and (2.19). Thus, we obtain

$$E(T) = \frac{1}{2} zV \frac{(1+A)XY}{1+AXY} . \quad (4.2)$$

Given the internal energy (4.2), we can calculate the specific heat *C(T)*

$$C(T) = \frac{dE(T)}{dT} , \quad (4.3)$$

and the Helmholtz free energy

$$F(T) = E(T^*) - T \int_{T^*}^{T} \frac{E(\tilde{T}) - E(T^*)}{\tilde{T}^2} d\tilde{T} , \quad (4.4)$$

where the limit $T^* \to 0$ is understood; the entropy $S(T) = (E-F)T$ follows immediately. The thermal compressibility $\kappa^T$ can be computed as:

$$\kappa^T = \frac{1}{n^2} \frac{dn}{d\mu}. \quad (4.5)$$

In Fig. 2 we show the chemical potential µ as a function of the particle density *n* for different values of the temperature *T*. The chemical potential is an increasing function of *n*, showing that the inhomogeneous phase is thermodynamically stable. As required by the particle-hole symmetry, $\mu$ satisfies the relation $\mu(1-n) = zV - \mu(n)$. At *T*=0, µ takes the following values

$$\mu = \begin{cases} 0 & 0 \leq n < 0.5 \\ 3V/2 & n = 0.5 \\ 3V & 0.5 < n \leq 1 . \end{cases}$$

In the region $0 \leq n < 0.5$ the repulsive intersite interaction disfavors the occupation of contiguous shells: only the even (odd) shells are occupied. The chemical potential takes the value $\mu = 0$ because there is no cost in energy to add one electron. At *n*=0.5 all even (odd) shells are occupied and all odd (even) shells are empty: that is, a checkerboard order is established. By increasing *n*, because of the Pauli principle, contiguous shells start to be occupied; in the region $0.5 < n \leq 1$ the cost in energy to add one electron is *zV*. The behavior at *n*=0.5 is shown in Fig. 3, where the particle densities $n_1$ and $n_2$ are plotted as a function of the temperature. We see that at *T*=0, $n_1 = 1$ and $n_2 = 0$, signaling the presence of a checkerboard structure.

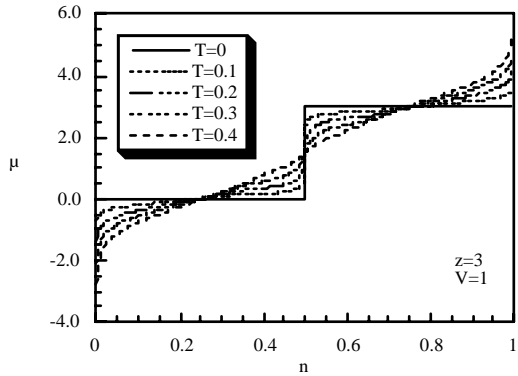 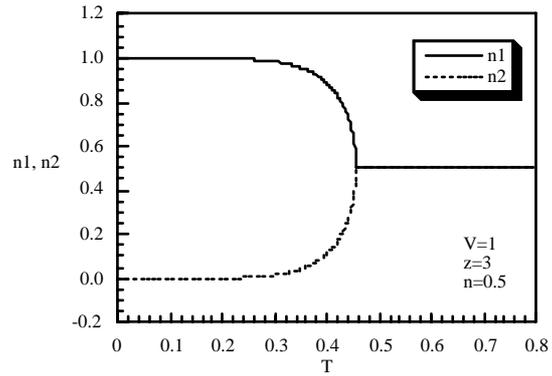

Fig. 2 The chemical potential $\mu$ is plotted as a function of the particle density $n$ at various temperatures.

Fig. 3 The particle densities on the even and odd shells, $n_1$ and $n_2$, are given as a function of the temperature at $n=0.5$.

In Fig. 4 we plot the particle densities $n_1$ and $n_2$ as function of the total particle density $n$ at several values of the temperature $T$. The regions inside the closed curves denote the inhomogeneous phase. When the temperature increases the area inside the curves shrinks and, at the critical temperature, there is a transition to the homogeneous disordered phase characterized by $n = n_1 = n_2$ and described by the straight line in Fig. 4.

The difference in the free energy $\Delta F = F_{\text{hom}} - F_{\text{inhom}}$ between the homogeneous and inhomogeneous phases is shown in Fig. 5 as a function of the particle density at various temperatures. We see that the inhomogeneous phase, when present, is always energetically favored.

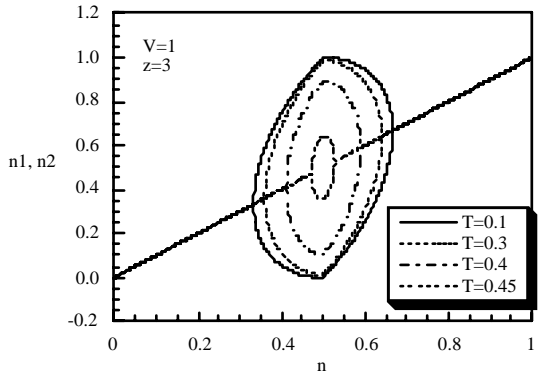 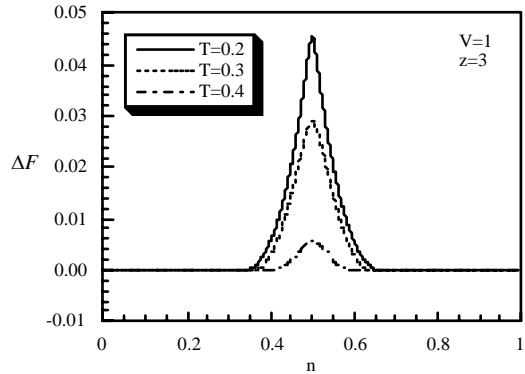

Fig. 4 The particle densities $n_1$ and $n_2$, as functions of the total density $n$ at various temperatures.

Fig. 5 The free energy difference $\Delta F = F_{\text{hom}} - F_{\text{inhom}}$ is plotted as a function of the particle density at various temperatures.

In Fig. 6 we report the behavior of the specific heat $C$ as a function of the temperature for $n=0.4$ and $n=0.5$. Owing to the particle-hole symmetry, the specific heat is invariant under the transformation $n \to 1-n$. We observe a jump in correspondence of the critical temperature, a clear signature of a second order phase transition.

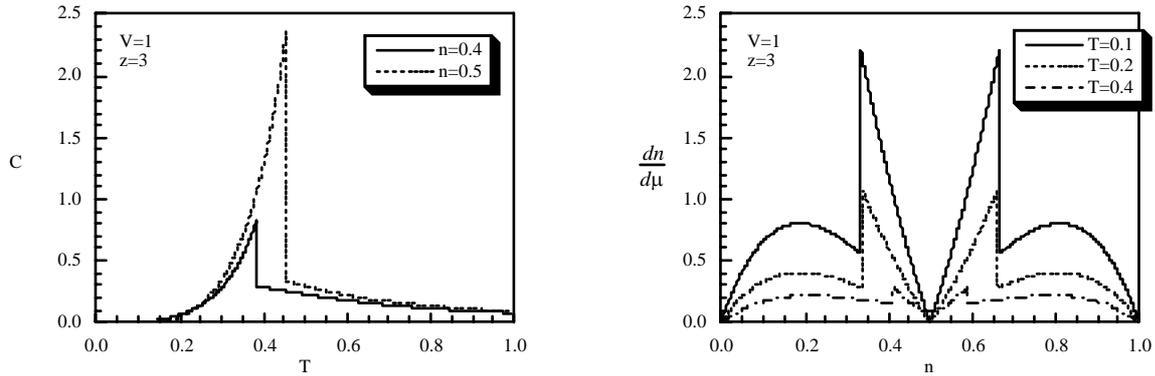

Fig. 6 The specific heat *C* is plotted as a function of the temperature *T* at *n*=0.4, 0.5.
Fig. 7 The derivative of the particle density with respect to the chemical potential is plotted as a function of *n* for *T*=0.1, 0.2, 0.4.

In Fig. 7 we study the thermal compressibility; we report the behavior of the derivative of the particle density with respect to the chemical potential as a function of *n* for some values of the temperature. We observe a jump in correspondence of the critical values of *n*, where the inhomogeneous phase establishes, in agreement with the phase diagram shown in Fig. 1. By increasing *T*, the height of the peaks decreases and the corresponding position moves towards *n*=0.5. At zero temperature the compressibility vanishes at *n*=0.5, where the checkerboard phase is observed.

Finally, the entropy *S* as a function of the particle density for various values of the temperature is shown in Fig. 8, and its temperature dependence is reported in Fig. 9 for *n*=0.4 and *n*=0.5. As expected, the entropy is lower in the region of *n* where a charge ordered state is observed. As a function of the temperature, the entropy exhibits a drastic change in correspondence of the critical temperature, in agreement with the results shown in Fig. 6 for the specific heat. This is another signature of a second order phase transition.

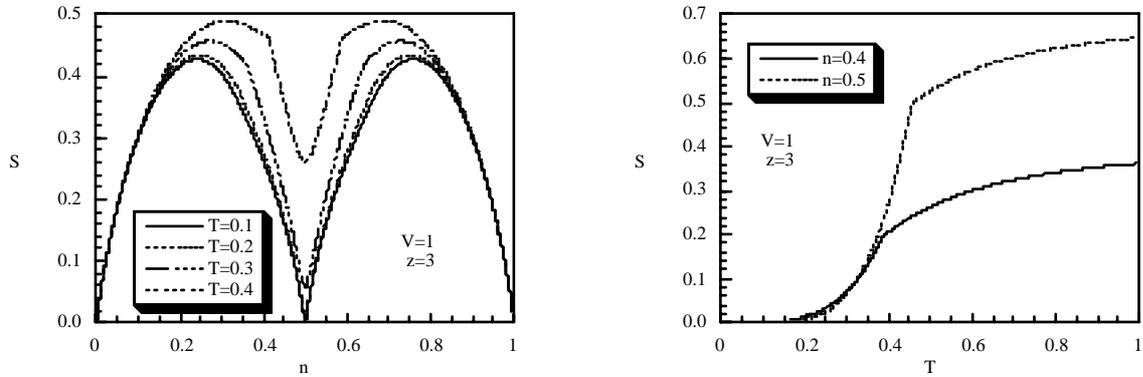

Figure 8: The entropy *S* is plotted as a function the particle density for various values of the temperature.
Figure 9: The entropy *S* is plotted against the temperature for *n*=0.4 and *n*=0.5.

## 5. Concluding Remarks

In this paper we have obtained the phase diagram at finite temperature of a system of spinless fermions, localized on the sites of the Bethe lattice, interacting through a repulsive nearest-neighbor potential. We have investigated the possibility of a spontaneous breakdown of the translational invariance and we have found that, below a critical temperature, a transition towards a thermodynamically stable charge ordered state occurs. The obtained phase diagram is well supported by the behavior of different physical quantities. Indeed, both the specific heat and the thermal compressibility present a jump in correspondence of the critical temperature and of the critical values of *n*, respectively. Furthermore, the behavior of the free energy and of the entropy clearly indicates that the inhomogeneous phase is always energetically favored.